\documentclass[twocolumn,showpacs,preprintnumbers,amsmath,amssymb,prb]{revtex4}

\usepackage{graphicx}
\usepackage{dcolumn}
\usepackage{bm}

\def\pd{\phantom{\dagger}}
\begin{document}


\title{Strong vs. Weak Coupling Duality and Coupling Dependence of the Kondo Temperature
in the Two-Channel Kondo Model}

\author{Christian Kolf}
\author{Johann Kroha}
\affiliation{Physikalisches Institut, Universit\"at Bonn, Nussallee 12, D-53115 Bonn, Germany}
\date{\today}

\begin{abstract}
We perform numerical renormalization group (NRG) as well as analytical 
calculations for the two-channel Kondo model to obtain the dependence of the 
Kondo temperature $T_K$ on the dimensionless (bare) spin exchange 
coupling $g$ over the complete parameter range from $g\ll 1$ to $g\gg 1$.
We show that there exists a duality between the regimes of small and large
coupling. It is unique for the two-channel model and enables 
a mapping between the strong and the weak coupling cases via the
identification $g\leftrightarrow 3/(2g)$, implying an exponential dependence 
of $T_K$ on $1/g$ and $g$, respectively, in the two regimes. 
This agrees quantitatively with our NRG calculations where we extract $T_K(g)$
over the complete parameter range and obtain 
a nonmonotonic $T_K(g)$ dependence, strongly peaked at the 2CK fixed point
coupling $g^*$. These results 
may be relevant for resolving the long-standing puzzle within the 2CK 
interpretation of certain random defect systems, why no broad distribution 
of $T_K$ is observed in those systems.
\end{abstract}


\maketitle

\section{Introduction}
The Kondo effect\cite{kondo64} is a paradigm for strong electronic correlations
in metals, induced by resonant quantum spin scattering of electrons at the
Fermi energy from local defects with spin $S$. The generalization of the 
problem to the case of $M$ equivalent conduction electron channels, the 
multi-channel Kondo problem, has attracted much attention ever since it
has been introduced by Nozi\`{e}res and Blandin \cite{nozieres80} in 1980.
While for a channel number $M=2S$ the impurity spin is exactly 
compensated by the conduction electron spins below the Kondo temperature
$T_K$, corresponding to 
 a spin singlet strong coupling fixed point 
with Fermi liquid behavior \cite{hewson93}, 
they showed that for $M>2S$ both the weak and the strong 
coupling fixed points are unstable, and hence a stable 
intermediate coupling fixed point was conjectured. 
It corresponds to an overcompensation of the
impurity spin at low temperatures due to the simultaneous 
screening by each channel,  
implying a nonvanishing zero-point entropy and non-Fermi liquid 
behavior. In the following we will focus our discussion
on the spin $S=1/2$ two-channel Kondo (2CK) effect.
The anomalous behavior of various thermodynamic quantities near the 
2CK fixed point has been worked out theoretically 
using the Bethe ansatz \cite{andrei84,tsvelik84}, a Majorana fermion 
representation of the problem,~\cite{tsvelik} 
and conformal field theory.~\cite{affleck,affleck91,affleck93} 
Early on, the two-level-system (TLS) model of atomic defects embedded 
in a metallic host was put forward by Zawadowski and Vladar \cite{zawa,zawadowski}
as a physical realization of 2CK defects, where the internal TLS degree of
freedom takes the role of the Kondo spin (pseudospin) and the magnetic conduction
electron spin serves as the conserved channel degree of freedom. 
However, it was shown thereafter 
that, unfortunately, within this model the 2CK fixed point cannot be reached 
because of the instability of the 2CK fixed point with respect to 
external perturbations: Within this model the TLS tunneling attempt frequency
sets the band cutoff for the 2CK effect, since band electrons at higher
energies instantaneously screen the tunneling defect without pseudospin flip. 
This turns out to prevent $T_K$ to be greater than the tunneling-induced
level splitting of the TLS.~\cite{altshuler} It remains to be seen if this 
problem can be overcome by a recently proposed modified TLS model, 
\cite{zarand05} where the 2CK fixed point may be stabilized by an
additional resonance enhancement of the conduction electron density of 
states (DOS).

On the experimental side, signatures consistent with the 2CK effect have 
been observed in both, certain bulk heavy 
fermion compounds \cite{maple,cox87,cichorek05} and in mesoscopic defect 
structures.\cite{ralph92,ralph95} The existence of TLS fluctuators in nanoconstrictions
has been established by various experiments.~\cite{keijsers96,heinrich05} 
One of the best-studied case of 2CK signatures is perhaps the zero-bias conductance 
anomaly observed by Ralph {\it et al.} in nanoscopic point contacts of
simple metals.~\cite{ralph92,ralph95,vdelft98}  
A scaling analysis of the differential conductance of these contacts 
\cite{ralph94,hettler} and systematic parameter variations lend strong support 
to the 2CK hypothesis. However, the 2CK interpretation of these data has
remained controversial \cite{altshuler_comment,ralph_reply} 
due to the lack of an established microscopic model for the physical
realization of the 2CK defects. 
See Ref. \onlinecite{kozub} for an alternative, statistical explanation of the
zero bias anomaly.   
Most recently, 2CK behavior seems to have been realized by systematically 
tuning a semiconductor quantum dot system into the 2CK 
regime,~\cite{goldhaber06}  as proposed theoretically in Ref.~\onlinecite{oreg}.

One of the problems with the 2CK interpretation of the anomalies in disordered, mesoscopic 
nanoconstrictions is the fact that within this interpretation these systems
exhibit a sharp value of the Kondo temperature $T_K$, while one expects
a broad distribution of the pseudospin flip coupling $J$ due to the random nature of
the 2CK defects. In fact, for single-channel Kondo impurities in
nanoconstrictions the observed behavior \cite{yanson95} has consistently been
explained \cite{zarand96} in terms of a broad $T_K$ distribution, induced by mesoscopic
fluctuations of the local DOS.

In the present paper we make a contribution to the resolution
of this puzzle. We compute the dependence of $T_K(J)$ on $J$ within the generic,
symmetric 2CK model, covering the complete range from small to large $J$.
Since the 2CK fixed point is at an intermediate coupling $J^*$, one expects
that for $J=J^*$ the 2CK regime extends in energy up to the band 
cutoff $D$,~\cite{zawadowski,florens04}
i.e. for the 2CK case $T_K(J)$ should have a maximum at $J=J^*$ with $T_K(J^*)\simeq D$.  
In Section II we define the model and, following the ideas of 
Nozi\`{e}res and Blandin,~\cite{nozieres80} establish a duality between the large
$J$ and the small $J$ region which makes it possible to give analytical
expressions for $T_K(J)$ in both regimes. Details of this calculation can be
seen in the Appendix.
In addition, we compute the complete dependence $T_K(J)$ using the NRG, as
explained in Section III. 
The results are presented in Section IV, which are in quantitative agreement
with the analytic expressions of section II and indicate a strongly peaked
dependence of $T_K$ on $J$. The conclusions and possible consequences for   
the 2CK interpretation of anomalies in nanoconstrictions are drawn in Section V.

\section{Duality of the 2CK weak and strong coupling regimes}
We consider the isotropic 2CK Hamiltonian,
\begin{equation}
  H_{2CK} = \sum_{k\alpha\sigma} \varepsilon_k c_{k\alpha\sigma}^{\dagger}
  c_{k\alpha\sigma}^{\phantom{\dagger}} + \frac{J}{2}\sum_{\alpha\sigma\sigma'} c_{0\alpha\sigma}^{\dagger} \vec{\sigma}_{\sigma\sigma'}c_{0\alpha\sigma}^{\phantom{\dagger}}\cdot\vec{S}
\label{Ham_2CK}
\end{equation}
where $c_{k\alpha\sigma}^{\dagger}$ are the usual creation 
operators for electrons in channel number $\alpha=\pm 1$ 
with momentum $k$ and spin $\sigma=\uparrow, \downarrow$.  
$c_{0\alpha\sigma}^{\dagger}=\sum_{k}c_{k\alpha\sigma}^{\dagger}$ is the 
creation operator for an electron at the impurity site, 
$\vec{\sigma}_{\sigma\sigma'}$ the vector of Pauli matrices 
and $\vec{S}$ the impurity spin operator of size $1/2$. 
The exchange
coupling $J>0$ is taken to be antiferromagnetic. We define the 
dimensionless coupling $g=\rho_0J$, where $\rho_0=1/2D$ is the DOS at the
Fermi level. Throughout this paper, all energy scales and
coupling constants are given in units of the band cutoff $D$.

In the weak coupling regime, $g\ll 1$, the crossover scale to the 
2CK non-Fermi liquid behavior can be obtained by perturbative analysis 
in $g$. It is well known as the weak coupling Kondo 
temperature and reads, including subleading logarithmic 
corrections,~\cite{hewson93}
\begin{equation}
T_K^{(wc)}\simeq D
{\rm e}^{-\frac{1}{2g}+\ln{(2g)}+\mathcal{O}(g)}, \qquad g\ll 1
\label{TK_weak}
\end{equation}

Turning now to the strong coupling regime, $g\gg 1$, it is convenient to 
consider the Hamiltonian in site representation, 
\begin{eqnarray*}
\sum_{k\alpha\sigma} \varepsilon_k c_{k\alpha\sigma}^{\dagger}
  c_{k\alpha\sigma}^{\phantom{\dagger}} = t
  \sum_{\langle i,j\rangle\alpha\sigma} 
  c_{i\alpha\sigma}^{\dagger}c_{j\alpha\sigma}^{\phantom{\dagger}}\ ,
\end{eqnarray*}
where $i$ is the site index and an infinite, one-dimensional lattice with 
a nearest neighbor hopping amplitude $t$ is assumed without loss of
generality. In the limit $g\to \infty$ the kinetic energy in the Hamiltonian
Eq.~(\ref{Ham_2CK}) is negligible, and we have,
\begin{equation}
  H^{(sc)} =
  \frac{J}{2}\sum_{\alpha\sigma\sigma'} 
  c_{0\alpha\sigma}^{\dagger} \vec{\sigma}_{\sigma\sigma'}c_{0\alpha\sigma'}^{\phantom{\dagger}}\cdot\vec{S}.
\label{Ham_sc}
\end{equation}
The mapping of the strong coupling regime of the 2CK model (\ref{Ham_2CK}) 
onto a weak coupling problem proceedes in two steps. We first represent the
Hamiltonian (\ref{Ham_2CK}) in the basis of low-lying eigenstates of the
strong coupling Hamiltonian (\ref{Ham_sc}), which will be of the type
of a generalized Anderson impurity model. Then we project this model 
in the low-energy regime onto an effective weak coupling 2CK model.  

The ground states of this strong coupling Hamiltonian (\ref{Ham_sc}) are  
3-body states comprised of one electron in each of the two channels,
located at the impurity site and antiferromagnetically coupled 
to the impurity spin. These states are easily calculated as     
\begin{eqnarray}
|\Psi_{\uparrow}^0\rangle = \frac{1}{\sqrt{6}}\left(
 2\,|\uparrow\Downarrow\uparrow\rangle - |\downarrow\Uparrow\uparrow\rangle -
 |\uparrow\Uparrow\downarrow\rangle\right) & = & F_{\uparrow}^{\dagger}|vac\rangle\label{gs1}\\
|\Psi_{\downarrow}^0\rangle = \frac{1}{\sqrt{6}}\left(
 2\,|\downarrow\Uparrow\downarrow\rangle - |\uparrow\Downarrow\downarrow\rangle -
 |\downarrow\Downarrow\uparrow\rangle\right) & = & F_{\downarrow}^{\dagger}|vac\rangle\label{gs2}
\end{eqnarray}
and have the energy $E_0=-J$, 
$H^{(sc)}|\Psi_{\uparrow(\downarrow)}^0\rangle = -J|\Psi_{\uparrow(\downarrow)}^0\rangle$.
In the Dirac ket notation above the thick arrow represents the impurity spin, 
while the first and the third (thin) arrow describes the
conduction electron spin in the $\alpha=+1$ and $\alpha=-1$ channel,
respectively. For later use we have also defined fermionic 
operators $F_{\sigma}^{\dagger}$ which create these states
out of the vacuum $|vac\rangle$ (i.e. the free Fermi sea without impurity).
Note that the ground states 
cannot simply be product states of 2-particle singlets, but necessarily 
contain triplet admixtures, a frustration effect implied by the
quantum nature of the Hamiltoniam (\ref{Ham_sc}). The  
degeneracy of the $|\Psi_{\uparrow(\downarrow)}^0\rangle$ is the reason why 
the 2CK model remains nontrivial even in the strong coupling limit, 
in contrast to the single-channel Kondo model. 
The next excited eigenstates of Eq.~(\ref{Ham_sc}) are the 2-body
singlet and triplet states $|\Psi_{sm\alpha}\rangle$,
\begin{eqnarray}
|\Psi_{001}\rangle & = & \frac{1}{\sqrt{2}}\left( |\uparrow\Downarrow 0\rangle - |\downarrow\Uparrow
  0\rangle\right) \label{ex3} \\
|\Psi_{101}\rangle & = & 
\frac{1}{\sqrt{2}}\left( |\uparrow\Downarrow 0\rangle + |\downarrow\Uparrow
  0\rangle\right) \label{ex4}\\
|\Psi_{111}\rangle & = &  |\uparrow\Uparrow 0\rangle \label{ex5}\\
|\Psi_{1-11}\rangle & = & |\downarrow\Downarrow 0\rangle , 
\label{ex6}
\end{eqnarray}
and analogous definitions for the $\alpha=-1$ channel.
In the above notation, $s=0,1$ denotes the total spin, 
$m=0,\pm 1$ its $z$-component and $\alpha =\pm 1$ the 
occupied conduction channel of the 2-body state. The energies of these
states with respect to Eq.\ (\ref{Ham_sc}) are 
$E_{00\alpha}=-\frac{3}{4} J$ and  $E_{1m\alpha}=+\frac{1}{4} J$,   
respectively. 
Switching on the hopping $t$ removes an electron from the 3-body 
states Eqs.\ (\ref{gs1}) and puts it onto a site $i\neq 0$ in the
conduction band. In this way, 8 states are generated which can be
expressed in terms of the strong coupling eigenstates 
Eqs.~(\ref{ex3})-(\ref{ex6}), see Appendix. It follows that in the
strong coupling eigenbasis Eqs.~(\ref{gs1})-(\ref{ex6}) the 2CK
Hamiltonian (\ref{Ham_2CK}) takes the form of a generalized 
2-channel Anderson impurity model, Eq.~(\ref{Anderson}), 
where the $|\Psi_{\sigma}^0\rangle$
play the role of the occupied and the $|\Psi_{sm\alpha}\rangle$
the role of the unoccupied impurity. By a straight-forward 
Schrieffer-Wolff transormation~\cite{swtrafo} for low 
energies, $\omega \ll J$, this Hamiltonian is 
projected onto the 2CK model
\begin{equation}
H^{(sc)}_{2CK} = 
t \sum_{<ij>, i,j\neq 0\ \alpha\sigma}c_{i\alpha\sigma}^{\dagger}
c_{j\alpha\sigma}^{\phantom{\dagger}}
+ \frac{\tilde J}{2}\sum_{\alpha\sigma\sigma'} c_{0\alpha\sigma}^{\dagger}
\vec{\sigma}_{\sigma\sigma'}c_{0\alpha\sigma'}^{\phantom{\dagger}}
\cdot\vec{\tilde S},
\label{Ham_2CKsc}
\end{equation}
where $\vec {\tilde S} = \sum _{\tau\tau'} F_{\tau}^{\dagger} 
\vec{\sigma}_{\tau\tau'} F_{\tau'}^{\phantom{\dagger}}$ is the 
spin operator of the strong coupling compound, and
$\tilde J = (1/\gamma) (4t)^2/J$, with $\gamma = 30/46 \approx 2/3$,
is the effective spin flip coupling in the strong coupling regime 
(see the Appendix for a detailed derivation). Using,
like in our NRG calculation of the following section, a
flat DOS of $\rho_0=1/4t$, the dimensionless coupling reads,
$\tilde g = \rho_0 \tilde J$.  
Following Eq.~(\ref{TK_weak}), the Kondo temperature is consequently 
given in the strong coupling regime by,
\begin{equation}
T_K^{(sc)}\simeq D
{\rm e}^{-\frac{1}{2} \gamma g - \ln{\left(\frac{\gamma}{2}
    g\right)}+\mathcal{O}(1/g)}, \qquad g \gg 1
\label{TK_strong}
\end{equation}
Comparison of Eq.~(\ref{TK_strong}) with Eq.~(\ref{TK_weak}) exhibits the duality of the 2CK model in the weak and strong coupling limits via the identification
\begin{equation}
\frac{1}{\rho_0J}\leftrightarrow \gamma \rho_0J \ .
\label{duality}
\end{equation}

\begin{figure}[t]
\begin{center}
\includegraphics[width=8.5cm,clip]{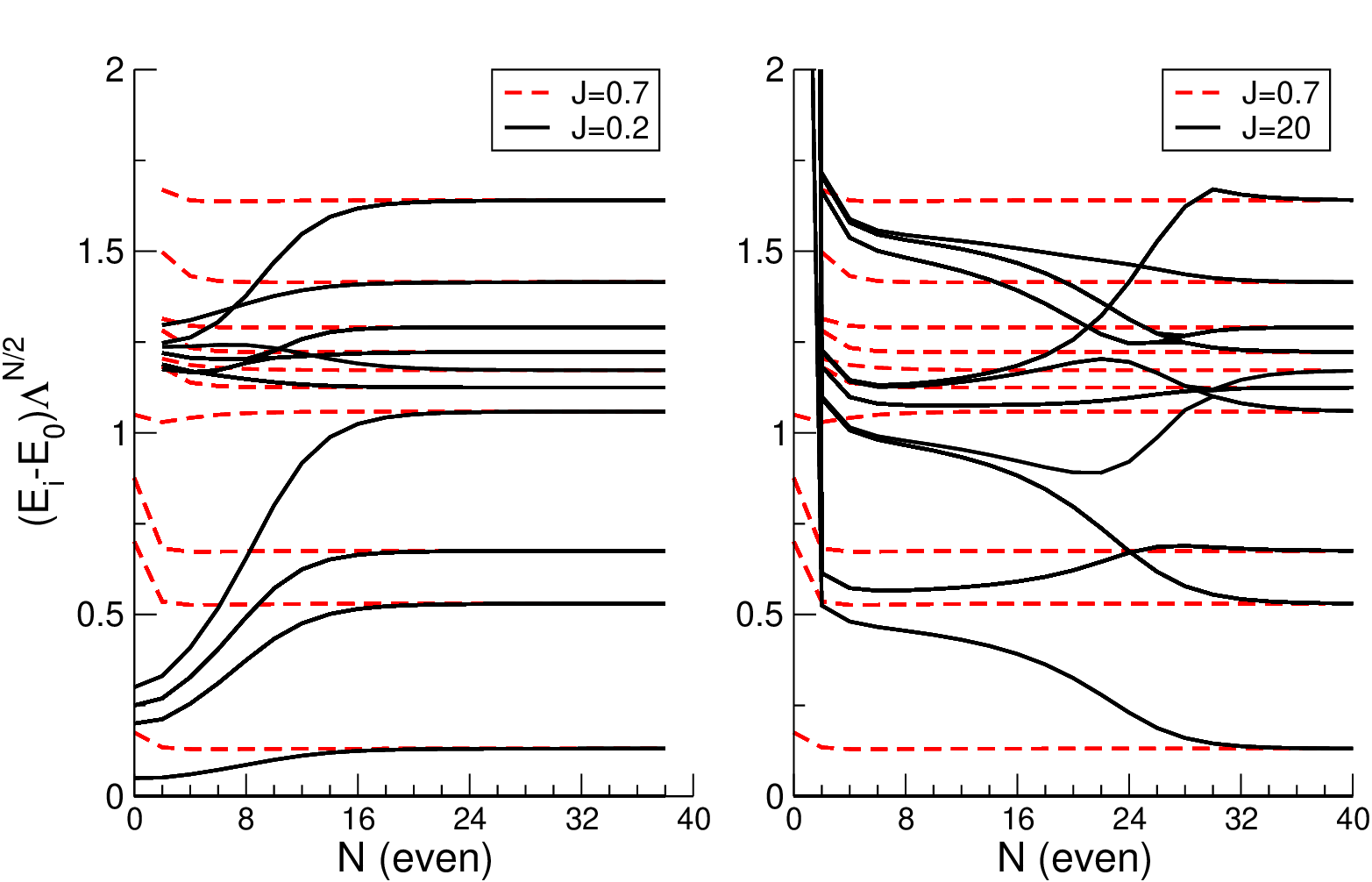}
\caption{\label{fig1}(Color online) Lowest energy levels of the isotropic 2CK model
as a function of the number of NRG iterations $N$ ($N$ even) 
for different initial couplings $J$ and $\Lambda=4$ with 900 states kept at
each iteration. As a guide to the eye, the levels obtained in the different NRG 
iterations $N$ are connected by straight lines.
Independently of initial weak-coupling ($J=0.2$),
intermediate coupling ($J=0.7$) or strong-coupling ($J=20$) strengths, 
the same fixed point spectrum is reached. For odd numer of iterations $N$
a non-equidistant fixed point spectrum is obtained as well (not shown).}
\end{center}
\end{figure}

\section{NRG treatment and results}
For the numerical solution of the 2CK problem we developed an efficient NRG
code, following Wilsons's original algorithm.~\cite{wilson}
Since for the two-channel model
the Hilbert space dimension grows particularly fast  
with the number $N$ of NRG iterations, i.e. as $16^N$, the use of conservation laws is
essentual to reduce the Hamiltonian to block structure.
The $M$-channel spin-$\frac{1}{2}$ Kondo model has a full symmetry group of
$SU(2)_{spin}\times Sp(M)$, where $Sp(M)$ is the symplectic group.~\cite{affleck92} In the two-channel case ($M=2$), the only 
decompositions into invariant subgroups of $Sp(2)$ are (i) $SU(2)\times U(1)$,
corresponding to channel and charge conservation, and (ii)
$SU(2)\times SU(2)$, corresponding to a separate axial charge
conservation, used in the work of Pang and Cox.~\cite{pang}
In our implementation of the NRG for the 2CK model, we have chosen to use the decomposition (i), where we use the charge $Q$,
$z$-component of the total (Kondo) spin $S_{tot}^z$ and the $z$-component of the channel spin as labels for the many particle
states only, corresponding to the following conserved operators,
\begin{eqnarray*}
\hat{Q} & = &\sum_{n=0,\alpha,\sigma}^{\infty}\left[f_{n\alpha\sigma}^{\dag}f_{n\alpha\sigma}^{\pd}-\frac{1}{2}\right]\\
\hat{S}_{tot}^z & = &\sum_{n=0,\alpha,\sigma}^{\infty}\sigma f_{n\alpha\sigma}^{\dag}f_{n\alpha\sigma}^{\pd} + S^z\\
  \hat{S}_{ch}^z & = &\frac{1}{2}\sum_{n=0,\alpha,\sigma}^{\infty} \alpha f_{n\alpha\sigma}^{\dagger}f_{n\alpha\sigma} \ .
\end{eqnarray*}
Thus we exploit only the $U(1)$ subgroups of the full
$SU(2)$ spin and channel symmetries, respectively. 
Accordingly, our code effectively uses a $U(1)\times U(1)\times U(1)$ symmetry. 
This turned out to be an optimal compromize between computatinal efficiency and
programming clarity.
The Hamiltonians are diagonalized in each irreducible subspace $|Q,
S_{tot}^z,S_{ch}^z\rangle$ and about 900 states were sufficient to be retained at each NRG
iteration.
\begin{figure}[t]
\begin{center}
\includegraphics[width=8.5cm,clip]{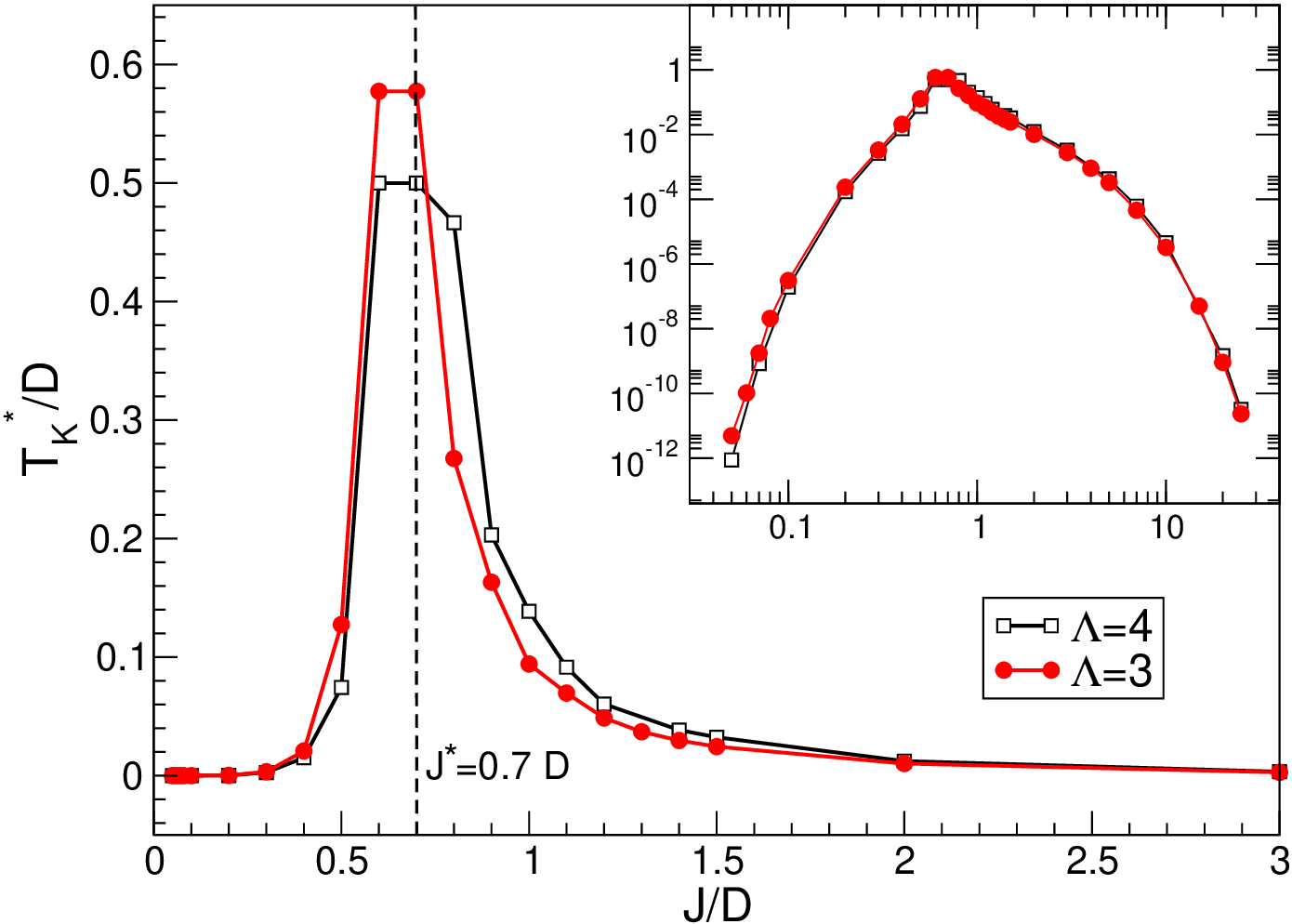}
\caption{\label{fig2}(Color online) Dependence of the Kondo temperature $T_K$
on the bare coupling strength $J$ (both in units of $D$), as
determined by NRG for $\Lambda =3$ and $\Lambda =4$.
The inset shows $T_K$ on a logarithmic scale.}
\end{center}
\end{figure}
After each NRG iteration the Hamiltonian is rescaled by the parameter $\sqrt{\Lambda}$,
$\Lambda >1$.~\cite{wilson} The correct convergence of the NRG procedure was checked 
by comparing the results obtained with two different $\Lambda$-values, 
$\Lambda =3$ and $\Lambda =4$. It yielded excellent quantitative agreement, as seen below
in Fig.~\ref{fig3}.

We have solved the isotropic 2CK model for a wide range of bare spin 
couplings $J$ in order to determine the $J$-dependence of $T_K$. 
Typical flow diagrams of the energy eigenvalues are shown in Fig.~\ref{fig1},
exhibiting non-equidistant level spacings characteristic for the non-Fermi liquid 
fixed point.~\cite{zawadowski} 
The fixed point coupling $J^*$ is characterized by the fact that,
when the inital coupling is $J=J^*$, the energy eigenvalues settle 
immediately (after 1 or 2 iterations) to their fixed point values 
(red dashed curves in Fig.~\ref{fig1}). It is thus identified as $J^*\approx 0.7 D$
in agreement with Ref.~\onlinecite{pang}. Following standard procedures, the
Kondo temperature $T_K$ can be determined as the energy scale either where the 
energy flow diagrams have an inflection point or where the first excited 
energy level has reached its fixed point value within, e.g., 10 percent. 
Both definitions of this crossover scale are equivalent up to a constant prefactor, 
as seen for the weak coupling region in Fig.~\ref{fig3}. 
Since, however, in the strong coupling region, $J>J^*$, the 
complexity of the level flow makes it difficult to identify a single inflection
point (see Fig.\ref{fig1}), we adopt the second definition. 

\begin{figure}[t]
\begin{center}
\includegraphics[width=8.5cm,clip]{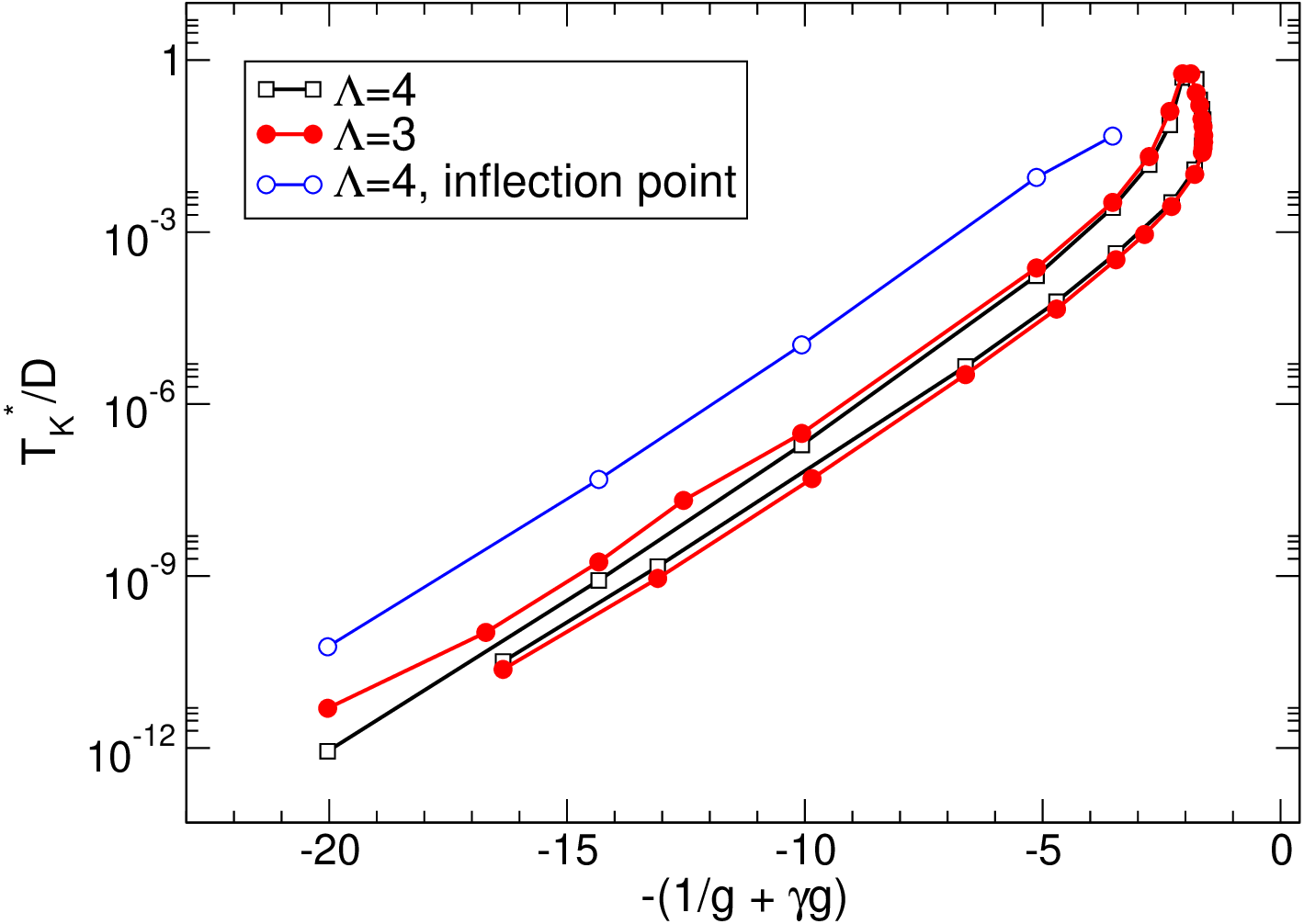}
\caption{\label{fig3}(Color online) 
The Kondo temperature $T_K$ is shown on a logarithmic scale 
versus the parameter $-(\frac{1}{\rho_0J}+\gamma\rho_0J)$, 
$\gamma =30/46\approx 2/3$.  
The upper branch of the curves corresponds to the weak coupling,
the lower branch to the strong coupling regime. The results for 
$T_K$ obtained from the ``inflection point method'' (see text)
in the weak coupling regime ($J<J^*$) are
shown for comparison and differ only by a constant prefactor.}
\end{center}
\end{figure}

Our results for the dependence of $T_K$ on the bare Kondo coupling $J$ are 
shown in Fig.~\ref{fig2}. It shows a strong peak 
at around $J=0.7 D$, as expected.
The results for the two discretizations considered, $\Lambda=3$, $\Lambda=4$, 
show no significant 
differences. The deviations in the intermediate coupling regime around the peak 
maximum in Fig.~\ref{fig2} arise from the difficulty to determine the exact $T_K^*$ 
when the crossover happens at the very beginning of the NRG iterations where the 
energy resolution is low. The behavior of $T_K^*$ is examined over nearly three 
decades of $J$ and extends over 
more than 10 decades in $T_K^*$, as illustrated in the inset of Fig.~\ref{fig2}.

The $J$-dependence of $T_K$ can be further analysed by plotting it in Fig.~\ref{fig3} 
versus the parameter $-(1/g + \gamma g)$. 
It shows the exponential behavior of $T_K$ as a function of $1/J$ in the 
weak coupling limit $1/J\rightarrow \infty$ and as a function of $J$ in the 
strong coupling limit $1/J\rightarrow \infty$, with logarithmic corrections 
towards the intermediate coupling regime, in agreement with Eqs.~\ref{TK_weak} and
\ref{TK_strong}. Note that the strong coupling and the weak coupling branches 
in Fig.~\ref{fig3} are parallel to each other, i.e. the NRG quantitaively 
confirms the analytical value of the parameter $\gamma=30/46$. The shift of the two 
branches can be traced back to the fact the strong coupling Anderson impurity 
model, Eq.~(\ref{Anderson}), produces
higher-order logarithmic corrections which are different from those of the 
weak coupling model, Eq.~(\ref{Ham_2CK}) and which
are, thus, not included in the effective low-energy 2CK model, Eq.~(\ref{Ham_2CKsc}). 
\section{Conclusion}
The two-channel Kondo model exhibits for low energies 
a duality between the regions of
weak and strong bare Kondo coupling $J$. This results from the fact that
in both limits, $J\to 0$ and $J\to \infty$ the ground state is doubly
degenerate. While for $J\to 0$ it is the decoupled impurity spin doublet,
for $J\to \infty$ it is a doubly degenerate quantum frustrated
3-body state, comprised of the impurity spin and the
conduction electron spins located at the impurity site in each of the two
channels. We have shown that, hence, the complete strong coupling
behavior can be obtained from the solution in the weak coupling regime
via the identification of the dimensionless coupling, 
$\gamma g \to 1/g$, where $\gamma = 30/46\approx 2/3$. These results have
been confirmed quantitatively by the exact numerical renormalization group
solution of the problem. 

As a result, the dependence of the Kondo temperature $T_K$ on the $J$ is
strongly peaked at the two-channel Kondo fixed point coupling, 
$J=J^*\approx 0.7$, and decays exponentially both for small and for
large couplings. The maximum is of the order of the band cutoff, 
$T_K(J^*)\approx D$, with non-Fermi liquid behavior for all energies
below $T_K$. 

We conjecture that this could be the reason why in experimental 
conductance anomalies of nanoconstrictions with two-channel Kondo 
signatures~\cite{ralph92,ralph95,vdelft98} no broad distribution 
of $T_K$ is observed: The band cutoff and hence $T_K(J^*)$ in 
two-channel Kondo systems can be provided by a decoherence scale of the
order of a few Kelvin.~\cite{altshuler} This would mean that,
even if there is a broad distribution of bare couplings, 
only for those couplings sufficiently close to $J^*$ the non-Fermi
liquid behavior would extend up to sufficiently high energies to
be observable. However, more detailed calculations as well as a 
detailed microscopic model for the two-channel Kondo defects will be
required to substantiate this conjecture.

\section*{Acknowledgments}
We would like to thank R.~Bulla, F.~B.~Anders and T. A. Costi for fruitful discussions
concerning the NRG. This research is supported by the DFG through the 
Collaborative Research Center SFB 608 and by grant No. KR1726/1.

\appendix*
\section{Details on the duality analysis}
Destruction of an electron from the 3-particle compound ground states 
(Eqs.~\ref{gs1}), (\ref{gs2}) in channel $\alpha=\pm 1$ 
generates the (unnormalized) states,
\begin{widetext}
\begin{eqnarray}
\begin{array}{rclcl}
c_{0\uparrow 1}|\Psi_{\uparrow}^0\rangle & = &
\frac{2}{\sqrt{6}}|0\Downarrow\uparrow\rangle -
\frac{1}{\sqrt{6}}|0\Uparrow\downarrow\rangle & = & \frac{\sqrt{3}}{2}|\Psi_{00-1}\rangle + \frac{1}{\sqrt{12}}|\Psi_{10-1}\rangle\\
c_{0\downarrow 1}|\Psi_{\uparrow}^0\rangle & = &
-\frac{1}{\sqrt{6}}|0\Uparrow\uparrow\rangle & = &-\frac{1}{\sqrt{6}}|\Psi_{11-1}\rangle\\
c_{0\uparrow -1}|\Psi_{\uparrow}^0\rangle & = &
\frac{2}{\sqrt{6}}|\uparrow\Downarrow 0\rangle -
\frac{1}{\sqrt{6}}|\downarrow\Uparrow 0\rangle & = & \frac{\sqrt{3}}{2}|\Psi_{001}\rangle + \frac{1}{\sqrt{12}}|\Psi_{101}\rangle\\
c_{0\downarrow -1}|\Psi_{\uparrow}^0\rangle & = &
-\frac{1}{\sqrt{6}}|\Uparrow\uparrow 0\rangle & = &-\frac{1}{\sqrt{6}}|\Psi_{111}\rangle\\
c_{0\uparrow 1}|\Psi_{\downarrow}^0\rangle & = &
-\frac{1}{\sqrt{6}}|0\Downarrow\downarrow \rangle & = &-\frac{1}{\sqrt{6}}|\Psi_{1-1-1}\rangle\\
c_{0\downarrow 1}|\Psi_{\downarrow}^0\rangle & = &
\frac{2}{\sqrt{6}}|0\Uparrow\downarrow\rangle -
\frac{1}{\sqrt{6}}|0\Downarrow\uparrow\rangle & = & -\frac{\sqrt{3}}{2}|\Psi_{00-1}\rangle + \frac{1}{\sqrt{12}}|\Psi_{10-1}\rangle\\
c_{0\uparrow -1}|\Psi_{\downarrow}^0\rangle & = &
-\frac{1}{\sqrt{6}}|\downarrow\Downarrow 0\rangle & = & -\frac{1}{\sqrt{6}}|\Psi_{1-11}\rangle\\
c_{0\downarrow -1}|\Psi_{\downarrow}^0\rangle & = &
\frac{2}{\sqrt{6}}|\downarrow\Uparrow 0\rangle -
\frac{1}{\sqrt{6}}|\uparrow\Downarrow 0\rangle & = &
-\frac{\sqrt{3}}{2}|\Psi_{001}\rangle + \frac{1}{\sqrt{12}}|\Psi_{101}\rangle
\end{array} 
\label{destruct}
\end{eqnarray}
which can be expressed in terms of the strong coupling singlet/triplet
eigenstates Eqs.~(\ref{ex3})-(\ref{ex6}) as indicated. 
We define bosonic creation operators for the latter states,
\begin{equation}
|\Psi_{sm\alpha}\rangle = B^{\dagger}_{sm\bar\alpha}|vac\rangle \ ,
\end{equation} 
which transform with respect to the channel SU(2) group according to the
adjoint representation, i.e. $\bar\alpha =-\alpha$.
Together with the fermionic operators of Eqs.~(\ref{gs1}), (\ref{gs2}) they
satisfy the constraint 
\begin{equation}
\hat Q = \sum_{\sigma} F^{\dagger}_{\sigma} F^{\phantom{\dagger}}_{\sigma} +
\sum_{sm\bar\alpha} B^{\dagger}_{sm\bar \alpha} B^{\phantom{\dagger}}_{sm\bar \alpha} 
=1\ ,
\end{equation} 
an expression of the uniqueness of the strong coupling basis states.
In the strong coupling basis, using Eq.~(\ref{destruct}) the 2CK 
Hamiltonian (\ref{Ham_2CK}) then takes form of a generalized
two-channel Anderson impurity model in one dimension,
\begin{eqnarray}
H = && t\sum_{\langle i,j\rangle \ i,j\neq 0}\sum_{\alpha\sigma}
c_{i\alpha\sigma}^{\dagger}c_{j\alpha\sigma}^{\phantom{\dagger}} 
 - J \sum_{\sigma} F_{\sigma}^{\dagger}F_{\sigma}^{\phantom{\dagger}} -
\frac{3}{4}J \sum_{\alpha}
B_{00\bar\alpha}^{\dagger}B_{00\bar\alpha}^{\phantom{\dagger}} + \frac{1}{4}J
\sum_{m=0,\pm 1\alpha} B_{1m\bar\alpha}^{\dagger}B_{1m\bar\alpha}^{\phantom{\dagger}}\nonumber\\
&& + \ t \sum_{i=\pm 1} \sum_{\alpha} \left[
  \frac{\sqrt{3}}{2}c_{i\alpha\uparrow}^{\dagger}B_{00\bar{\alpha}}^{\dagger}F_{\uparrow}^{\phantom{\dagger}} 
  -\frac{1}{\sqrt{6}}c_{i\alpha\downarrow}^{\dagger}B_{1+1\bar{\alpha}}^{\dagger}F_{\uparrow}^{\phantom{\dagger}}
  +\frac{1}{\sqrt{12}}c_{i\alpha\uparrow}^{\dagger}B_{10\bar{\alpha}}^{\dagger}F_{\uparrow}^{\phantom{\dagger}} \right.\nonumber\\
&&\hspace*{1.75cm}
\left.-\frac{\sqrt{3}}{2}c_{i\alpha\downarrow}^{\dagger}B_{00\bar{\alpha}}^{\dagger}F_{\downarrow}^{\phantom{\dagger}} 
      -\frac{1}{\sqrt{6}}c_{i\alpha\uparrow}^{\dagger}B_{1-1\bar{\alpha}}^{\dagger}F_{\downarrow}^{\phantom{\dagger}}
      +\frac{1}{\sqrt{12}}c_{i\alpha\downarrow}^{\dagger}B_{10\bar{\alpha}}^{\dagger}F_{\downarrow}^{\phantom{\dagger}}+ H.c.\right] \ ,
\label{Anderson}
\end{eqnarray}
where $V=2t$ plays the role of the band-impurity hybridization and the factor
2 arises from the fact that there is hopping from the impurity site $0$ to the
two sites $i=\pm 1$. By means of a Schrieffer-Wolff transformation~\cite{swtrafo} this Hamiltonian maps for energies $\omega\ll J$ onto the
effective 2CK model Eq.~(\ref{Ham_2CKsc}), where potential scattering terms
have been neglected. Since only the intermediate (bosonic) states with $m=0$
contribute to an effective Kondo spin flip (i.e. only products of the  
1st and the 4th term and of the 3rd and the 6th term of the hybridization part
in Eq.~(\ref{Anderson})), the effective spin flip coupling, as defined through
Eq.~(\ref{Ham_2CKsc}) reads,
\begin{equation*}
\tilde J =2\frac{4t^2}{J}\left(
\frac{\frac{3}{4}}{1-\frac{3}{4}}+\frac{\frac{1}{12}}{1+\frac{1}{4}}\right) =
\frac{1}{\gamma}\frac{(4t)^2}{J} \ ,
\end{equation*}
where $\gamma = 30/46 \approx 2/3$.
\end{widetext}
\bibliographystyle{apsrev} 
\bibliography{2CKpaper}
\end{document}